\documentclass{emulateapj}
\usepackage{CJK}
\usepackage{longtable,lscape}
\usepackage{natbib}
\usepackage[dvips]{color}
\usepackage{multirow}
\newcommand{\HII}{H\,{\sc ii}}
\newcommand{\HI}{H\,{\sc i}}

\newcommand{\LZ}{{\it L-Z}}
\newcommand{\MZ}{{\it M-Z}}

\begin{document}
%\begin{CJK*}{UTF8}{gbsn}
%\shorttitle{Chemical Properties of BCDs}
%\shortauthors{Zhao, Gao \& Gu}
\submitted{To appear in The Astrophysical Journal}
\title{A Study on the Chemical Properties of Blue Compact Dwarf Galaxies}
\author{Yinghe Zhao\altaffilmark{1, 2 ,3}, Yu Gao\altaffilmark{1,2} and Qiusheng Gu\altaffilmark{4, 3}}
\altaffiltext{1}{Purple Mountain Observatory, Chinese Academy of Sciences, Nanjing 210008, China; yhzhao@pmo.ac.cn}
\altaffiltext{2}{Key Laboratory of Radio Astronomy, Chinese Academy of Sciences, Nanjing 210008, China}
\altaffiltext{3}{Key Laboratory of Modern Astronomy and Astrophysics (Nanjing University), Ministry of Education, Nanjing 210093, China}
\altaffiltext{4}{School of Astronomy and Space Science, Nanjing University, Nanjing 210093, China}
%\date{Received:~ Accepted:~}
\begin{abstract}
In this paper, we report our studies on the gaseous and chemical properties of a relatively large sample (53 members) of blue compact dwarf galaxies (BCDs). The results of correlations among the oxygen abundance, stellar mass, gas mass, baryonic mass, and gas fraction are present both for E- and I-type BCDs, which are classified according to Loose \& Thuan (1985) and show elliptical and irregular outer haloes, respectively. These correlations of I-type BCDs show similar slopes to those of E-type ones. However, in general, E-type BCDs are more gas-poor and metal-rich than I-type ones at a given baryonic mass. Based on these results, we suggest that E-type BCDs, at least a part of them, and I-type ones might be likely at different evolutionary phases and/or having different progenitors. Our investigation of the correlation between oxygen abundance and gas fraction shows that BCDs appear to have not evolved as isolated systems, but to have experienced some gas flows and/or mergers.  
\end{abstract}

\section{Introduction}
Dwarf galaxies are by far the most numerous galaxies in the universe (e.g. Mateo 1998) and play a crucial role in our understanding of the formation and evolution of galaxies. They are proposed to be the building-blocks in hierarchical scenarios of galaxy formation (Kauffmann et al. 1993). Dwarf galaxies are typically metal-poor (e.g. Mateo 1998), and have low stellar masses. The low metallicities of these systems make them having similar conditions of high-redshift galaxies. Therefore, dwarf galaxies at the local universe can be used as a proxy to reveal the physical properties of some galaxies at the early universe.

Based on their morphological properties and luminosity, dwarf galaxies can be classified into different types, e.g., blue compact dwarfs (BCDs), dwarf ellipticals (dEs), dwarf irregulars (dIs), and etc. BCDs, which are the focus of the current paper, have dramatically different properties compared to normal dwarf galaxies (Zwicky 1966; Gil de Paz et al. 2003, hereafter G03). They are small, gas-rich (H\,{\sc i} mass fraction typically higher than 30\%; please see Section 2.3 and 3.1) extragalactic objects. Being metal-poor ($0.03 Z_\odot\lesssim Z \lesssim Z_\odot$, here $12+({\rm O/H})_\odot=8.69$ (Allende Prieto et al. 2001; Asplund et al. 2009); e.g. Terlevich et al. 1991; Izotov et al. 1994; Izotov \& Thuan 1999; Hunter \& Hoffman 1999), these galaxies are allowing us to study the star formation and chemical enrichment in a near-primordial environment. As such, they are important to our understanding of galaxy formation and evolution.

BCDs are spectroscopically characterized by a faint, blue optical continuum accompanied, in most cases, by intense and narrow emission lines, due to the intense star formation activity distributed in one to several star-forming regions (Cair\'os et al. 2001a). dEs have smooth visual appearances and low gas content, and they represent the bright, gas-poor dwarf galaxies which have little star formation (e.g. Binggeli 1994). By contrast, dIs are characterized by a lack of spiral arms  and an irregular visual appearance (see review by Hunter 1997). They have low dust content (e.g., Calzetti 2001) and large gas reservoirs (see, e.g., Hoffman et al. 1996). Central surface brightness increases with increasing luminosity in all of these types of dwarf galaxies (see recent review of Tolstoy et al. 2009), thus it seems natural to consider a possible evolutionary relationship between BCDs, dIs and dEs (e.g. Thuan 1985; Ferguson \& Binggeli 1994; Papaderos et al. 1996). 

However, the relation between BCDs, dIs, and dEs/dSphs remains unclear. Current knowledge bearing on this question comes mainly from analysis of the photometric properties (such as structure, surface brightness profile, and color) of these three types of dwarf galaxies (e.g. Thuan 1985; James 1994; Papaderos et al. 1996; Noeske et al. 2000; Gil de Paz \& Madore 2005; Vaduvescu et al. 2006). The results from different authors seem to be contradictory. For example, Thuan (1985) suggests that BCDs and dIs are the same type of galaxy, differing only in their present star formation rate, and they can evolve into dEs. However, James (1994) obtains an opposite result that BCDs are probably dEs undergoing starburst activities, while dIs are a fundamentally distinct population from them. The results in Papaderos et al. (1996) show that BCDs have a brighter central surface brightness and a smaller exponential scale length than dIs and dEs at a given $B$-band luminosity. However, Gil de Paz \& Madore (2005) claim that at least $\sim$15\% of BCDs in their sample have structural properties compatible with being dEs (also see Papaderos et al. 1996) experiencing a burst of star formation, whereas Vaduvescu et al. (2006) conclude that BCDs have similar structures to dIs. Recently, Vaduvescu \& McCall (2008) suggest that there are strong connections among BCDs, dIs, and dEs since these galaxies lie precisely on a fundamental plane defined by the $K_{\rm s}$-band absolute magnitude, the central surface brightness, and the neutral hydrogen line width (also see McCall et al. 2012).

Some other attempts to address this issue rely on the chemical evolution of dwarf galaxies. Richer \& McCall (1995) show that [O/Fe] for a number of field dEs are larger than those in field dIs at similar luminosities and conclude that there is unlikely an evolutionary connection between dEs and dIs. Vaduvescu et al. (2007; hereafter VMR07) argue that the evolution of BCDs has been similar to that of dIs since they have the same metallicity-gas fraction correlations. However, the sample in VMR07 is too small (14 BCDs), and therefore might suffer large uncertainties. Moreover, the correlation between metallicity and the gas fraction for BCDs and dIs shows large dispersions (see Figure 10 in VMR07). 

In our previous works, we have studied the luminosity (or mass)-metallicity (\LZ) relation (Zhao et al. 2010; hereafter ZGG10), stellar population and star formation (Zhao et al. 2011; hereafter ZGG11) for a sample of BCDs. Our results show that BCDs are old systems currently experiencing starbursts, and follow a well-defined \LZ\ relation. Based on these two works and G03, along with \HI\ data compiled from literature (see reference in Table 1), we can further investigate the evolution of gaseous and chemical properties of BCDs. In this paper, we will explore correlations among different galactic parameters, i.e. oxygen abundance, stellar mass, gas mass, baryonic mass, and gas mass fraction. We also compare these relations for morphologically different types of BCDs and dIs. Our investigations on such correlations may shed some light on the answer to the question of whether there is an evolutionary connection between BCDs, dIs and dEs.

The paper is organized as following: Section 2 describes the sample of BCDs and the data. In Section 3, we first  calculate the gas masses using the compiled \HI\ fluxes and derive the stellar masses, and then present several relationships involving stellar mass, gas mass and oxygen abundance. We also do some comparisons between dIs and BCDs in this section. Based on the results shown in Section 3, we discuss possible evolutionary connections between different types of dwarf galaxies, and gas flows in BCDs, in Section 4. A brief summary is given in the last section.

\section{Sample and Data}

\begin{deluxetable*}{lcccccccccccc}
\tablenum{1}
\tablecaption{Properties of our sample BCDs}
\tablewidth{0pc}
\tabletypesize{\scriptsize}
\tablehead{
&\colhead{DM$^{\rm a}$}&\colhead{$B^{\rm a}$}&\colhead{$R^{\rm a}$}&&&&\colhead{$\log M_{{\rm{gas}}}$}&\colhead{$\log M_\star$}&&&\\
\colhead{Galaxy}&\colhead{(mag)}&\colhead{(mag)}&\colhead{(mag)}&\colhead{$T^{\rm a}$}&\colhead{12+log(O/H)}&\colhead{M}&\colhead{ ($M_\odot$)} & \colhead{($M_\odot$)} &\colhead{M\tablenotemark{b}}& \colhead{ $\mu$} & \colhead{$\log\log (1/\mu)$}&\colhead{Ref}\\
\colhead{(1)}&\colhead{(2)}&\colhead{(3)}&\colhead{(4)}&\colhead{(5)}&\colhead{(6)}&\colhead{(7)}&\colhead{(8)}&\colhead{(9)}&\colhead{(10)}&\colhead{(11)}&\colhead{(12)}&\colhead{(13)}}
\startdata
            Haro 2\dotfill & 31.67 & 13.39 & 12.87 & nE & $ 8.38\pm0.03$& $T_e$ & $ 8.87\pm0.03$ & 9.43 & 2 & $ 0.22\pm0.17$& $-0.18\pm0.14$& 1\\
            Haro 3\dotfill & 30.79 & 13.22 & 12.61 & iE & $ 8.46\pm0.10$& $T_e$ & $ 8.93\pm0.01$ & 8.92 & 2 & $ 0.51\pm0.24$& $-0.53\pm0.09$& 1\\
            Haro 4\dotfill & 29.73 & 15.59 & 15.44 & iI & $ 7.82\pm0.02$& $T_e$ & $ 7.51\pm0.02$ & 7.01 & 4 & $ 0.76\pm0.20$& $-0.92\pm0.05$& 1\\
            Haro 8\dotfill & 31.02 & 14.27 & 13.20 & nE & $ 8.35\pm0.04$& $T_e$ & $ 8.30\pm0.06$ & 8.54 & 2 & $ 0.37\pm0.24$& $-0.36\pm0.12$& 2\\
           Haro 14\dotfill & 30.73 & 13.65 & 12.91 & iE & $ 8.3\pm0.2$& N2 & $ 8.44\pm0.04$ & 8.93 & 1 & $ 0.24\pm0.18$& $-0.21\pm0.14$& 1\\
            I Zw 18\dotfill & 30.50 & 16.05 & 16.24 & i0 & $ 7.18\pm0.01$& $T_e$ & $ 8.05\pm0.09$ & 6.77 & 4 & $ 0.95\pm0.64$& $-1.65\pm0.13$& 2\\
           I Zw 123\dotfill & 30.34 & 15.42 & 14.85 & nE & $ 8.07\pm0.02$& $T_e$ & $ 7.95\pm0.07$ & 7.62 & 2 & $ 0.68\pm0.33$& $-0.78\pm0.09$& 1\\           
           II Zw 40\dotfill & 29.96 & 11.87 & 11.10 & iIM & $ 8.09\pm0.01$& $T_e$ & $ 8.76\pm0.03$ & 8.08 & 1 & $ 0.83\pm0.22$& $-1.09\pm0.05$& 2\\
           II Zw 70\dotfill & 31.36 & 14.84 & 14.29 & iI & $ 8.18\pm0.08$& $T_e$ & $ 8.68\pm0.01$ & 8.37 & 1 & $ 0.67\pm0.22$& $-0.77\pm0.06$& 1\\
           II Zw 71\dotfill & 31.36 & 14.45 & 13.54 & iI & $ 8.24\pm0.17$& $T_e$ & $ 9.13\pm0.01$ & 8.74 & 3 & $ 0.71\pm0.20$& $-0.83\pm0.05$& 1\\
           VII Zw 403\dotfill & 28.41 & 14.11 & 13.58 & iE & $ 7.70\pm0.01$& $T_e$ & $ 8.07\pm0.04$ & 7.31 & 1 & $ 0.85\pm0.23$& $-1.16\pm0.05$& 2\\
          Mrk 05\dotfill & 30.60 & 15.13 & 14.56 & iE & $ 8.06\pm0.04$& $T_e$ & $ 8.24\pm0.10$ & 8.12 & 1 & $ 0.57\pm0.33$& $-0.61\pm0.11$& 2\\
            Mrk 67\dotfill & 30.77 & 16.10 & 15.34 & nE & $ 8.08\pm0.08$& $T_e$ & $ 7.63\pm0.01$ & 7.75 & 2 & $ 0.43\pm0.24$& $-0.44\pm0.11$& 5\\
            Mrk 86\dotfill & 29.20 & 12.07 & 11.49 & iEr & $ 8.53\pm0.2$& N2 & $ 8.50\pm0.03$ & 8.98 & 1 & $ 0.25\pm0.18$& $-0.22\pm0.14$& 2\\
           Mrk 108\dotfill & 31.69 & 15.15 & 14.66 & iIM & $ 7.96\pm0.02$& $T_e$ & $10.06\pm0.03$ & 8.09 & 1 & $ 0.99\pm0.21$& $-2.34\pm0.04$& 2\\
           Mrk 178\dotfill & 28.11 & 14.15 & 13.60 & iE & $ 7.92\pm0.02$& $T_e$ & $ 7.28\pm0.06$ & 7.14 & 2 & $ 0.58\pm0.29$& $-0.62\pm0.09$& 2\\
           Mrk 209\dotfill & 28.82 & 14.15 & 13.94 & iE & $ 7.76\pm0.01$& $T_e$ & $ 8.04\pm0.03$ & 7.25 & 1 & $ 0.86\pm0.20$& $-1.18\pm0.04$& 2\\
           Mrk 324\dotfill & 32.01 & 15.17 & 14.60 & nE & $ 8.18\pm0.2$& N2 & $ 8.62\pm0.01$ & 8.76 & 1 & $ 0.42\pm0.24$& $-0.43\pm0.11$& 5\\
           Mrk 328\dotfill & 31.75 & 14.93 & 14.18 & nE & $ 8.66\pm0.01$& $T_e$ & $ 8.17\pm0.01$ & 8.89 & 1 & $ 0.16\pm0.13$& $-0.10\pm0.15$& 5\\
           Mrk 409\dotfill & 31.64 & 14.37 & 13.33 & iEr & $ 8.8\pm0.2$& N2 & $ 8.67\pm0.14$ & 9.58 & 3 & $ 0.11\pm0.10$& $-0.02\pm0.17$& 2\\
           Mrk 450\dotfill & 30.57 & 14.44 & 13.65 & iE & $ 8.12\pm0.02$& $T_e$ & $ 8.41\pm0.06$ & 8.48 & 1 & $ 0.46\pm0.26$& $-0.47\pm0.11$& 2\\
           Mrk 475\dotfill & 29.93 & 16.20 & 15.67 & nE & $ 7.93\pm0.02$& $T_e$ & $ 6.65\pm0.12$ & 7.44 & 2 & $ 0.14\pm0.12$& $-0.07\pm0.16$& 4\\
           Mrk 600\dotfill & 30.82 & 14.85 & 14.82 & iE & $ 7.94\pm0.06$& $T_e$ & $ 8.63\pm0.05$ & 7.81 & 1 & $ 0.87\pm0.29$& $-1.21\pm0.06$& 2\\
           Mrk 900\dotfill & 31.37 & 14.17 & 13.56 & nE & $ 8.07\pm0.03$& $T_e$ & $ 8.43\pm0.03$ & 8.96 & 1 & $ 0.23\pm0.17$& $-0.19\pm0.14$& 4\\
           Mrk 996\dotfill & 31.88 & 15.01 & 14.08 & nE & $ 8.2\pm0.2$& N2 & $ 8.37\pm0.07$ & 9.11 & 1 & $ 0.15\pm0.13$& $-0.09\pm0.16$& 4\\
          Mrk 1313\dotfill & 32.51 & 16.02 & 15.50 & i0 & $ 8.22\pm0.10$& $T_e$ & $ 8.74\pm0.01$ & 8.63 & 3 & $ 0.56\pm0.24$& $-0.60\pm0.08$& 5\\
          Mrk 1329\dotfill & 31.02 & 14.08 & 13.38 & iE & $ 8.26\pm0.02$& $T_e$ & $ 8.88\pm0.01$ & 8.64 & 1 & $ 0.64\pm0.23$& $-0.71\pm0.07$& 5\\
          Mrk1416\dotfill & 32.64 & 16.32 & 15.75 & iI & $ 7.84\pm0.02$& $T_e$ & $ 8.86\pm0.07$ & 8.34 & 4 & $ 0.77\pm0.36$& $-0.95\pm0.09$& 4\\
          Mrk1423\dotfill & 31.55 & 14.90 & 13.52 & nE & $ 8.5\pm0.2$& N2 & $ 8.05\pm0.11$ & 8.85 & 1 & $ 0.14\pm0.12$& $-0.06\pm0.16$& 4\\
          Mrk 1450\dotfill & 30.83 & 15.75 & 15.09 & nE & $ 7.96\pm0.02$& $T_e$ & $ 7.54\pm0.03$ & 7.82 & 2 & $ 0.34\pm0.22$& $-0.33\pm0.12$& 7\\
          Mrk 1480\dotfill & 32.18 & 16.17 & 15.56 & nE & $ 8.04\pm0.05$& $T_e$ & $ 8.83\pm0.01$ & 7.93 & 3 & $ 0.89\pm0.13$& $-1.28\pm0.03$& 7\\
          Mrk 1481\dotfill & 32.18 & 16.19 & 15.57 & iIM & $ 8.11\pm0.2$& N2 & $ 8.79\pm0.02$ & 8.46 & 3 & $ 0.68\pm0.22$& $-0.78\pm0.06$& 7\\
          NGC 2915\dotfill & 27.78 & 11.93 & 10.96 & iE & $ 8.4\pm0.2$& N2 & $ 8.64\pm0.03$ & 8.30 & 1 & $ 0.69\pm0.24$& $-0.79\pm0.06$& 3\\
          NGC 4861\dotfill & 30.50 & 12.68 & 11.91 & iIC & $ 8.00\pm0.01$& $T_e$ & $ 9.28\pm0.03$ & 8.53 & 1 & $ 0.85\pm0.19$& $-1.15\pm0.04$& 2\\
      SBS1054+504\dotfill & 31.52 & 16.08 & 15.46 & nE & $ 8.26\pm0.05$& $T_e$ & $ 7.66\pm0.03$ & 8.29 & 3 & $ 0.19\pm0.15$& $-0.14\pm0.15$& 7\\
      SBS1147+520\dotfill & 31.39 & 16.95 & 15.98 & nE & $ 8.2\pm0.2$& N2 & $ 7.82\pm0.01$ & 8.02 & 4 & $ 0.39\pm0.23$& $-0.38\pm0.11$& 7\\
      SBS1331+493\dotfill & 29.98 & 14.87 & 14.16 & iE & $ 7.78\pm0.02$& $T_e$ & $ 8.31\pm0.03$ & 8.00 & 4 & $ 0.67\pm0.24$& $-0.76\pm0.07$& 4\\
      SBS1415+437\dotfill & 30.02 & 15.43 & 14.77 & iIC & $ 7.59\pm0.01$& $T_e$ & $ 8.20\pm0.03$ & 7.74 & 4 & $ 0.74\pm0.23$& $-0.89\pm0.06$& 4\\
      SBS1428+457\dotfill & 32.74 & 15.42 & 14.67 & iI & $ 8.42\pm0.05$& $T_e$ & $ 9.21\pm0.01$ & 9.22 & 3 & $ 0.49\pm0.24$& $-0.51\pm0.09$& 7\\
          UGCA 184\dotfill & 31.81 & 15.99 & 15.80 & iI & $ 8.03\pm0.01$& $T_e$ & $ 8.57\pm0.18$ & 7.97 & 3 & $ 0.80\pm0.87$& $-1.01\pm0.21$& 4\\
          UGCA 412\dotfill & 33.07 & 15.55 & 14.62 & nE & $ 8.12\pm0.17$& $T_e$ & $ 8.38\pm0.11$ & 9.21 & 4 & $ 0.13\pm0.11$& $-0.05\pm0.16$& 4\\
            UM 133\dotfill & 31.91 & 15.41 & 14.51 & iIC & $ 7.69\pm0.02$& $T_e$ & $ 9.02\pm0.01$ & 8.77 & 4 & $ 0.64\pm0.22$& $-0.72\pm0.07$& 5\\
            UM 323\dotfill & 32.26 & 16.09 & 15.24 & iE & $ 7.96\pm0.04$& $T_e$ & $ 8.60\pm0.01$ & 8.49 & 3 & $ 0.56\pm0.24$& $-0.60\pm0.08$& 5\\
            UM 408\dotfill & 33.58 & 17.46 & 16.70 & nE & $ 7.74\pm0.05$& $T_e$ & $ 9.03\pm0.01$ & 8.46 & 4 & $ 0.79\pm0.17$& $-0.98\pm0.04$& 5\\
            UM 439\dotfill & 30.73 & 14.77 & 14.09 & iE & $ 8.08\pm0.03$& $T_e$ & $ 8.63\pm0.01$ & 7.83 & 3 & $ 0.86\pm0.12$& $-1.19\pm0.03$& 5\\
            UM 452\dotfill & 31.45 & 15.25 & 14.07 & nE & $ 8.27\pm0.05$& $T_e$ & $ 7.76\pm0.07$ & 8.80 & 3 & $ 0.08\pm0.08$& $ 0.03\pm0.17$& 5\\
            UM 455\dotfill & 33.64 & 17.02 & 16.26 & iI & $ 7.74\pm0.02$& $T_e$ & $ 9.21\pm0.03$ & 8.66 & 4 & $ 0.78\pm0.21$& $-0.97\pm0.05$& 5\\
            UM 491\dotfill & 32.20 & 15.54 & 14.95 & nE & $ 8.3\pm0.2$& N2 & $ 8.36\pm0.01$ & 8.68 & 3 & $ 0.32\pm0.21$& $-0.31\pm0.12$& 5\\
            UM 533\dotfill & 30.34 & 14.63 & 13.64 & iE & $ 8.55\pm0.29$& $T_e$ & $ 8.16\pm0.01$ & 8.55 & 4 & $ 0.29\pm0.20$& $-0.27\pm0.13$& 5\\
          VCC 0130\dotfill & 31.02 & 17.05 & 16.27 & iE & $ 8.3\pm0.2$& N2 & $ 7.94\pm0.05$ & 7.65 & 3 & $ 0.66\pm0.28$& $-0.75\pm0.08$& 6\\
          VCC 0459\dotfill & 31.02 & 14.95 & 14.13 & iI & $ 8.27\pm0.09$& $T_e$ & $ 8.30\pm0.01$ & 8.48 & 1 & $ 0.40\pm0.23$& $-0.40\pm0.11$& 6\\
          VCC 0655\dotfill & 31.02 & 13.32 & 12.12 & iEr & $ 8.7\pm0.2$& N2 & $ 7.99\pm0.03$ & 9.43 & 1 & $ 0.04\pm0.03$& $ 0.16\pm0.18$& 6\\
          VCC 0848\dotfill & 31.02 & 15.03 & 14.10 & iIM & $ 8.06\pm0.12$& $T_e$ & $ 8.93\pm0.01$ & 8.43 & 1 & $ 0.76\pm0.18$& $-0.93\pm0.04$& 6\\
\enddata
\tablecomments{
Column 1: galaxy name;  Column 2: distance moduli. Column 3--4: {\it B} and{\it R}-band apparent magnitudes, respectively. Column 5: morphological type. Column 6: oxygen abundances adopted from ZGG10; the error for the N2-based metallicity is the typical scatter comparing to those obtained with $T_e$ method. Column 7: method of determining oxygen abundance. Column 8: total gas mass. Column 9: stellar mass. Column 10: method of determining stellar mass (please see Section 3.1 for details). Column 11: gas fraction. Column 12: inverse gas fraction. Column 13: references for H\,{\sc i} data.}
\tablenotetext{a}{All data, except for morphological types of three galaxies (Mrk 005, Mrk 178 and Mrk 1329), are adopted from Gil de Paz et al. (2003), and all of the magnitudes have been corrected for the Galactic extinctions.}
\tablenotetext{b}{The stellar mass is derived with: 1=$K_s$-band data; 2=integrated spectra; 3=SDSS fiber spectra; 4=$R$-band data.}
\tablerefs{
(1): Gordon \& Gottesman 1981; (2): Thuan \& Martin 1981; (3): Huchtmeier \& Richter 1988; (4): Thuan et al. 1999; (5): Salzer et al. 2002; (6): Gavazzi et  al. 2005; (7): Huchtmeier et al. 2005.
}
\end{deluxetable*}

\subsection{The Sample}
Using a unified concept of BCD galaxy, G03 have compiled a BCD sample (including 105 members) from several exploratory studies. The galaxies in their sample were selected by putting forward a new set of quantitative classification criteria. This new set of criteria is a combination of the galaxy's color, morphology and luminosity. Briefly, a BCD galaxy has to fulfill the following three criteria (referring to {\it blue, compact} and {\it dwarf}, respectively): (1) $\mu_{B,\rm{peak}}-\mu_{R,\rm{peak}} < 1$, where $\mu_{B,\rm{peak}}$ and $\mu_{R,\rm{peak}}$ are the peak surface brightness of {\it B}- and {\it R}-band, respectively, (2) $\mu_{B,\rm{peak}} < 22$ mag arcsec$^{-2}$, and (3) the absolute magnitude $M_K > -21$ mag (see G03 for details). 

To study the chemical evolution of BCDs, we need to know their metallicities and gaseous properties. Based on the sample in G03, ZGG10 have selected a subsample of BCDs whose metallicities have been measured or having SDSS spectroscopic data. Therefore, the sample we used here is derived from ZGG10 and a galaxy is selected if its \HI\ data has been given in literature. The final sample contains 53 BCD galaxies, whose basic properties are listed in Table 1. The distance, $B$ and $R$ photometric data and morphological type (except three galaxies; see below) are all adopted from G03. Except for several galaxies whose distance had been determined by measuring the magnitude of the tip of the red giant branch (RGB), the majority were computed using the galactic standard of rest velocity of the galaxies assuming a Hubble constant, $H_0$, of 70 km s$^{-1}$ Mpc$^{-1}$ (Freedman et al. 2001). For the photometric data, the foreground Galactic extinctions have been corrected in G03, using $A_B$ values determined following Schlegel et al. (1998) and the Galactic extinction law of Cardelli et al. (1989) with $R_V=3.1$. We did not attempt to correct for internal extinction, since this is known to vary spatially in BCDs (e.g., Cannon et al. 2002; Garc\'{i}a-Lorenzo et al. 2008; Cair\'{o}s et al. 2010). The internal extinction obtained through Balmer lines and/or continuum fitting based on the aperture spectra can only be applied to a small part of a galaxy. In addition, it was derived only for about 40\% of the total sample.

The morphological types of these BCDs were classified according to Loose \& Thuan (1986). However, for Mrk 005 (Cair\'{o}s et al. 2001b; Amor\'{i}n et al. 2009), Mrk 178 (Schulte-Ladbeck et al. 2000; Noeske et al. 2003) and Mrk 1329 (Noeske et al. 2003; Vaduvescu et al. 2006), we adopt their morphological types as iE, based on the deeper optical and/or IR observations in the above references. Please refer to G03 for details of the aforementioned properties. In  the following, we refer to all nE and iE BCDs (showing an outer elliptical envelope) collectively as ``E-type" BCDs, and refer to all iI BCDs (showing an irregular outer halo; here including two i0 BCDs) as iI or ``I-type" BCDs unless otherwise stated.

The oxygen abundances are taken from ZGG10 (also see references therein). In the majority of the cases the oxygen abundances have been determined on the basis of electronic temperatures measured from the [O\,{\sc iii}]\,$\lambda4363$ line (i.e. $T_e$-method). In a few cases $T_e$-based abundances are not available, we adopt the N2-based oxygen abundances, which have been derived with the empirical N2-O/H relation proposed by Denicol\'o et al. (2002) (i.e. N2-method, where N2\,$\equiv\log$\,([N\,{\sc ii}]\,$\lambda$6584/H$_\alpha$)). As shown in ZGG10, for the BCD sample studied here, metallicities derived with the N2-method are generally consistent with those derived with the $T_e$ method, and have an accuracy of  $\sim0.2$ dex (also see Denicol\'o et al. 2002; Pettini \& Pagel 2004 and Salzer et al. 2005). Please see ZGG10 and references therein for details.

In Figure 1 we plot the integrated $(B-R)$ color and the absolute $B$ magnitude ($M_B$) distributions for the G03 sample and our subsamples using different histograms. We can see that ours have similar distributions of $(B-R)$ and $M_B$ to the G03 sample, except that our iI subsample contains a higher fraction of galaxies with $M_B\sim 16.5-17$ mag and a lower fraction of galaxies with $(B-R) > 1.0$. Therefore, our subsamples can be representative of the total sample of G03. However, please note that none of these samples are complete, although they covers a large range of galactic parameters.

\begin{figure}[t]
\centering
\includegraphics[bb=10 2 468 356,width=0.48\textwidth]{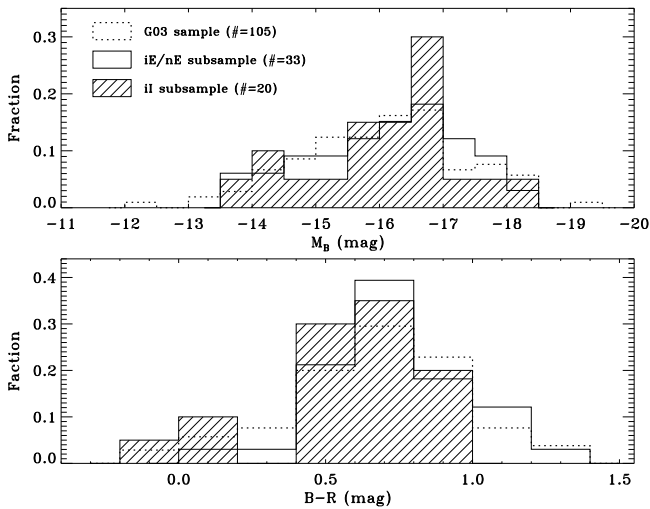}
\caption{Distributions of absolute $B$ magnitude (upper) and ($B-R$) color (bottom) for the three BCD samples. Our subsamples (hatched filled and solid open) show similar distributions to the G03 sample (dashed open).}
\label{Fig1}
\end{figure}

In order to explore whether their is an evolutionary link between BCDs and dIs by means of comparing their gaseous and chemical properties, we also include a sample of field dIs adopted from Lee et al. (2003) in our analysis. Lee et al. (2003) sample is compiled from the literature and consists of 22 dIs. The data for this dI sample, including metallicity, gas mass and stellar mass, are all adopted directly from Lee et al. (2003). The metallicities are determined by the direct method, and the gas masses are derived using the same method as described below. Whereas for the stellar mass, we will check the consistency in the next section.

\subsection{Stellar Mass}
The stellar mass, $M_\star$, of a galaxy, can be derived from its luminosity and mass-to-light ratio ($M_\star/L$), which can be estimated by comparing observed galaxy colors (e.g. Bell \& de Jong 2001), broadband spectral energy distributions (e.g. Walcher et al. 2008), or full optical spectra (e.g. Cid Fernandes et al. 2007) with predications from population synthesis models. 

For 21 galaxies in our sample, their stellar masses have been derived in ZGG11 by using optical spectra and stellar population synthesis code, STARLIGHT{\footnotemark\footnotetext{STARLIGHT \& SEAGal: http://www.starlight.ufsc.br/.}} (Cid Fernandes et al. 2005, 2007). During the fitting process, the Bruzual \& Charlot (2003) models were adopted, and the spectra of the simple stellar population library were computed with the Salpeter (1955) initial mass function (IMF), Padova 1994 models, and the STELIB library (Le Borgne et al. 2003), by limiting the metallicities to $0.4Z_\odot$ for this metal poor sample. Stellar masses obtained with this method are represented by 2 (using integrated spectra; Moustakas \& Kennicutt 2006) or 3 (using SDSS fiber spectra, and the fraction of the total galaxy luminosity in the SDSS {\it i} band inside the fiber had been used to correct for the aperture effect) in column 10 of Table 1. Please refer to ZGG11 for more details.

For the rest galaxies, we use the $M_\star/L$ ratio-color relations given by Bell \& de Jong (2001), which are derived by using the Bruzual \& Charlot (2003) model with a Salpeter IMF, to convert luminosity to stellar mass. Due to the ongoing starburst activities in BCDs, we prefer to use the near infrared luminosity ($L_{K_s}$) and $M_\star/L_{K_s}$ to calculate the stellar mass. This is because that the $K_s$-band luminosity is relatively less sensitive to recent star formation activity and dust extinction, and can well approximate the stellar mass of a galaxy. The $M_\star/L_{K_s}$ ratio-color relation, as given by Bell \& de Jong (2001), is applied to correct for the dependence of $M_\star/L_{K_s}$ on star formation history. For our metal-poor BCDs, we adopt the result derived with metallicity $Z=0.4Z_\odot$. In addition, 15 galaxies in our sample have both optical spectra and NIR photometric data. Therefore, we can compare the derived stellar masses. We find that the NIR-based stellar masses are systematically larger (0.11 dex with a dispersion of 0.23 dex) than spectrum-based ones. Therefore, we subtract this systematic difference from the NIR-based stellar masses. The results obtained with this method are also listed in Table 1, as represented by 1 in column 10. 

Eleven galaxies of our sample have neither optical spectra nor NIR photometry. For these galaxies we use $R$-band data and the $M_\star/L_{R}$-color relation in Bell \& de Jong (2001) to calculate their stellar masses. The $R$-band data might be more affected by recent star formation and extinction than NIR data. To check to what extent, we compare the $R$-band-based stellar masses with spectrum-based ones for 21 galaxies. We find that, in general, $R$-band-based stellar masses are 0.16 dex (with a dispersion of 0.27 dex) larger than spectrum-based ones. Hence,  a correction of -0.16 dex is applied to the $R$-band-based stellar masses, as represented by 4 in column 10 of Table 1.

As shown above, there exist some offsets between the stellar masses derived with different methods. The most possible reason causing this offset is the star formation histories (SFHs), as discussed in Gallazzi \& Bell (2009). For galaxies with busty (or continuous) SFHs, such as BCDs, the one color-based mass-to-light ratio is generally overestimated by 0.1-0.2 dex (Gallazzi \& Bell 2009), which agrees well with our results. The dust might also play a role in this discrepancy since we do not correct for the internal extinction for the optical data. However, dust extinction decreases the luminosity  and reddens the optical color. To first-order, these effects cancel out, and the estimated stellar mass of a galaxy is almost unaffected even if a moderate internal extinction of $E(B-V)=0.2$ mag is assumed (using the Cardelli et al. (1989) extinction law). The uncertainty in the derived stellar mass depends majorly on the uncertainties of models, galaxy evolution prescription, and SFHs, which are all of order of 0.1-0.2 dex or less (Bell \& de Jong 2001). Therefore, we adopt a statistical uncertainty of 0.3 dex for our results.

The stellar masses for the dI sample are derived with a two-component (``young" plus ``old") method (Lee et al. 2003). To check whether this method is consistent with ours, we calculated the $M_\star/L_B$ using the $(B-V)$ color and $M_\star/L_{B}$-color relation given by Bell \& de Jong (2001) for the dI sample, and then compared them with those derived with the two-component method. We found that the one-color based $M_\star/L_{B}$ is systematically larger by 0.14 dex, with a scatter of  0.11 dex, than that derived with the two-component method. This is generally in accordance with the result described above. Therefore, we conclude that the methods used to determine the stellar masses for the dI and BCD sample are consistent.

\subsection{Gas Mass and Fraction}
The total mass of atomic hydrogen, $M_{{\rm H\,I}}$,  in solar masses, can be derived from the \HI\ 21 cm observations by using the following equation (Roberts 1975; Roberts \& Haynes 1994):
\begin{equation}
M_{{\rm H\,I}} = (2.356\times10^5) F_{21} D^2,
\end{equation}
where $F_{21}$ is the 21 cm flux integral in Jy km s$^{-1}$ and $D$ is the distance in Mpc. To account for helium and other metals, the total gas mass in solar masses is given by
\begin{equation}
M_{\rm gas} = M_{\rm H\,I}/X.
\end{equation}
where X is the fraction of the gas mass in the form of hydrogen, here adopted to be 0.733 (Lee et al. 2003) in consideration of the similar range of metallicity and the consistency between the dI and BCD samples. In Table 1 we include the logarithm of the total gas mass for our BCD sample, calculated from $F_{21}$ (collected from literature; see references in Table 1) and the distance modulus included in the second column.

The gas fraction, which is a natural parameter for quantifying to what degree a galaxy has exhausted its available fuel supply, is defined as
\begin{equation}
\mu=\frac{M_{\rm gas}}{M_{\rm tot}}=\frac{M_{\rm gas}}{M_{\rm gas}+M_\star}
\end{equation}
where the total mass in baryons, $M_{\rm tot}$, is taken to be the mass of gas and stars. The gas fraction is a distance-independent quantity, because the gas mass and stellar mass are both derived from electromagnetic fluxes. Derived gas fractions for the sample of BCDs are summarized in Table 1. For the BCDs in our sample, about 75\% have their $\mu > 0.3$, and the median value of $\mu$ is 0.57.

\section{Results and Analysis}
\subsection{Correlations Involving Stellar Mass}
\subsubsection{Stellar Mass-Gas Mass Relation}
The total gas mass and stellar mass in our BCD sample both range from $\sim10^6$ to $\sim10^{10}$ $M_\odot$.  Figure 2 shows the distribution of total gas mass as a function of stellar mass for E-type and I-type BCDs. The solid line is a reference line for the case when the two qualities are the same. We find that, almost all I-type BCDs lie above the $1:1$ line, whereas more than half of the E-type BCDs just below this line, indicating that I-type BCDs tend to have larger gas masses than E-type BCDs at a given stellar mass. For I-type BCDs, except for the apparent merging system, Mrk 108, whose gas content is extraordinarily high, the gas mass well correlates with the stellar mass, namely galaxies with a larger stellar mass also contains higher gas masses. The E-type BCDs seem to have a similar $M_{\rm gas}-M_\star$ relation when $M_\star \lesssim 10^9$ $M_\odot$. While at the high-mass end ($M_\star \gtrsim 10^9 M_\odot$), galaxies tend to have lower gas contents comparing to the lower mass counterparts. As a comparison, in Figure 2 we also plot the data for field dIs which are adopted from Lee el al. (2003). In this plot, field dIs seem to distribute similarly to BCDs as a whole, but the $M_{\rm gas}-M_\star$ relation of field dIs has a smaller dispersion than that of BCDs.

\begin{figure}[t]
\centering
\includegraphics[bb=20 2 488 340, width=0.48\textwidth]{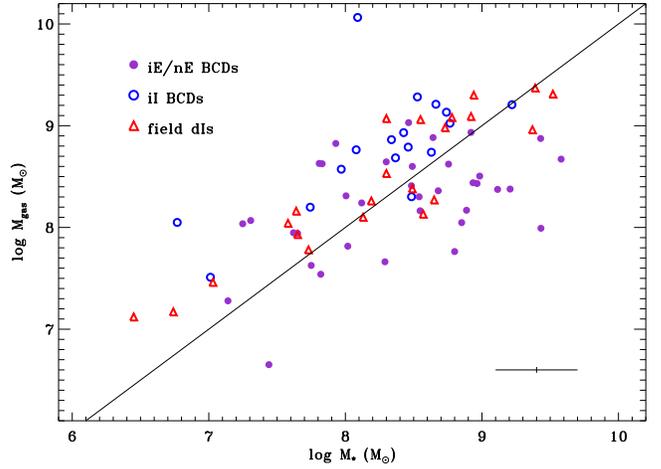}
\caption{Distribution of total gas mass for I- and E-type BCDs, and field dIs, as a function of stellar mass. The data for dIs are adopted from Lee et al. (2003). Solid line indicates $1:1$ ratio, and the cross in the bottom right gives the typical error bar of the BCD sample.}
\label{Fig2}
\end{figure}

Since we are interested in the potential for connections/differences between different types of dwarfs, it is informative to consider the gas fraction of a galaxy. In Figure 3 we plot the number distribution of gas fraction, $\log \mu$, in E- and I-type BCDs, and field dIs. As mentioned above, I-type BCDs are generally more gas-rich than E-type BCDs. The separation between these types of BCDs is more striking in Figure 3. The $\log\mu$ for E-type BCDs shows a wide distribution and ranges from $\sim -1.6$ to $\sim 0.0$, and has a median value of $-0.37$. However, the distribution of $\log\mu$ for I-type BCDs is much narrower, and looks similar to that for field dIs. I-type BCDs concentrate between $-0.2-0.0$ (14 out of 17), while relatively more field dIs have $-0.6\leqslant\log\mu\leqslant-0.2$. The median values of $\log\mu$ are $-0.12$ and $-0.19$, for I-type BCDs and field dIs, respectively.

To quantify the significance of the difference or similarity of the gas content among these different types of galaxies, we perform a nonparametric statistical test, the Kolmogorov-Smirnov (K-S) test. The K-S test for the I-type sample against the E-type sample rejects the null hypothesis (that these two data sets can be derived from a same parent population) at a $99.95\%$ significance level. On the other hand, the K-S tests give probabilities of $6.65\%$, $3.22\%$ and $26.64\%$, for the dIs vs I- and E-type BCDs, all BCDs as a whole, data sets, respectively. Therefore, we conclude that the gas fraction distributions of I- and E-type BCDs are much different, whereas the gas fraction distributions of dIs and BCDs seem to be indistinguishable. Nevertheless, we caution that such analysis is possibly limited by the small numbers of the samples.

\begin{figure}[t]
\centering
\includegraphics[bb=40 2 480 332,width=0.48\textwidth]{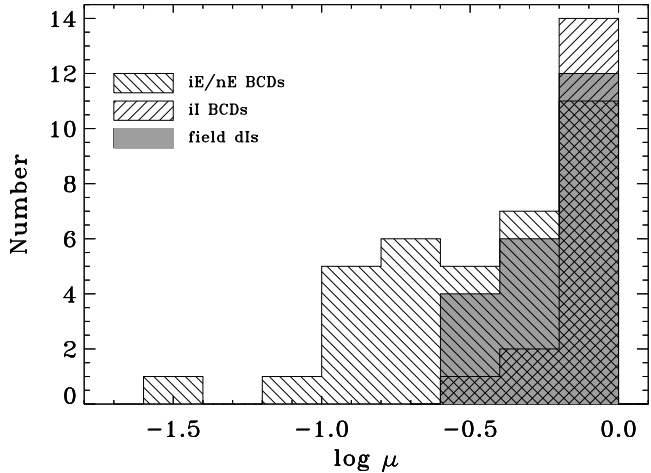}
\caption{Number distribution of gas fraction for different types of dwarf galaxies. I-type BCDs have a higher mean gas fraction than E-type ones and field dIs.}
\label{Fig3}
\end{figure}

\subsubsection{Stellar Mass-Gas Fraction Relation}
We show the gas fraction of different types of galaxies plotted against their stellar mass in Figure 4. At the low-mass end ($M_\star \lesssim 6.3\times10^8 M_\odot$), in general, the gas mass is roughly independent on the stellar mass (i.e. proportional to the baryonic mass). At higher masses, however, the relation exhibits a break and becomes considerably steeper since the gas has been more efficiently converted to stars in higher stellar mass galaxies. 

From Figure 4 we can see that most BCDs and field dIs distributes similarly in the $\log \mu$-$M_\star$ plot. However, at a given stellar mass, the gas fractions for E-type BCDs have a larger span and a lower mean value than those for I-type BCDs and field dIs, which confirms that E-type BCDs are generally more gas-poor.

\subsection{Correlations Involving Metallicities}
Star-forming dwarf galaxies are observed to be well mixed chemically throughout their high surface brightness components (Kobulnicky \& Skillman 1996, 1997), having no significant gradients or inhomogeneities in oxygen abundances measured from \HII\ regions located at different galactocentric radii (e.g. Lee \& Skillman 2004). 
In Figure 5 we show the metallicity distributions for E- and I-type BCDs and dIs. It seems that all of these different types of dwarfs show a double-peaked distribution of metallicity. However, E-type BCDs have a larger median oxygen abundance ($\sim8.2$) than I-type BCDs and dIs (both $\sim 8.0$). In the following we explore the chemical evolution for BCDs by investigating in detail the correlations between their oxygen abundance and gas mass, baryonic mass, and gas fraction.

\begin{figure}[t]
\centering
\includegraphics[bb=16 2 490 343,width=0.48\textwidth]{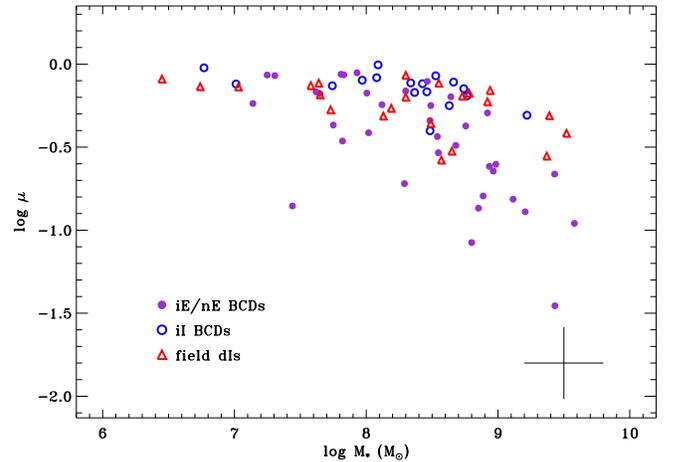}
\caption{Distribution of gas fraction ($\log \mu$) as a function of stellar mass for BCDs and dIs. The cross in the bottom right gives the typical error bar of the BCD sample. More massive systems have lower gas fractions.}
\label{Fig4}
\end{figure}

\subsubsection{Metallicity versus Gas Mass}
Figure 6 shows the metallicity as a function of the total gas mass. Albeit with large scatters, a galaxy with a greater gas mass is generally more abundant in metals. If we look into different types of galaxies, the scatter will reduce much.  A unweighted least-squares linear fit, using a geometrical mean functional relationship (Isobe et al. 1990), to the two BCD subsamples, gives, 
\[12+\log ({\rm O/H})=(3.65\pm0.84)+(0.55\pm1.42)\log M_{\rm gas},\]
\[12+\log ({\rm O/H})=(3.34\pm1.29)+(0.53\pm0.32)\log M_{\rm gas}.\]
for E- and I-type BCDs respectively. The fitted trends are shown as (purple) solid and (blue) dashed lines in Figure 6, and the rms deviations of the data from the relationships are 0.35 and 0.34 dex in (O/H), for E- and I-type BCDs, respectively. While for the dI sample, we obtain the following $({\rm {O/H}})-M_{\rm gas}$ relation
\[12+\log ({\rm O/H})=(4.53\pm0.36)+(0.40\pm0.10)\log M_{\rm gas}.\]
with a rms scatter of 0.26 dex in (O/H), and is shown with the (red) dot-dashed line in Figure 6.

\begin{figure}[t]
\centering
\includegraphics[bb=40 10 482 335,width=0.48\textwidth]{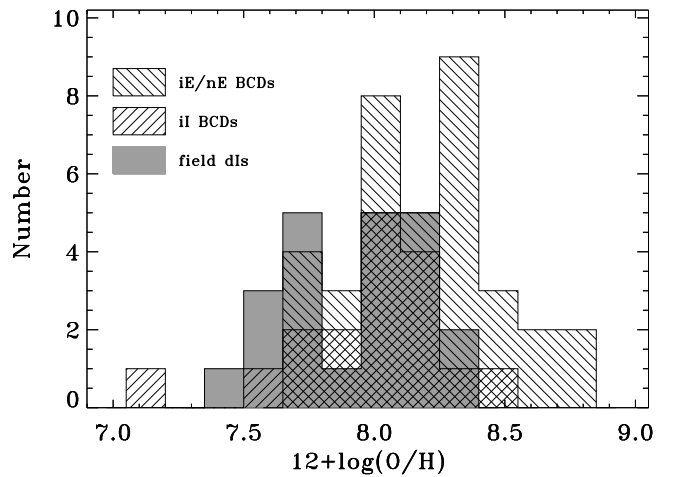}
\caption{Number distribution of metallicity for different types of dwarf galaxies. E-type BCDs have a higher mean metallicity than I-type ones and field dIs.}
\label{Fig5}
\end{figure}

\begin{figure}[t]
\centering
\includegraphics[bb=20 5 482 340,width=0.48\textwidth]{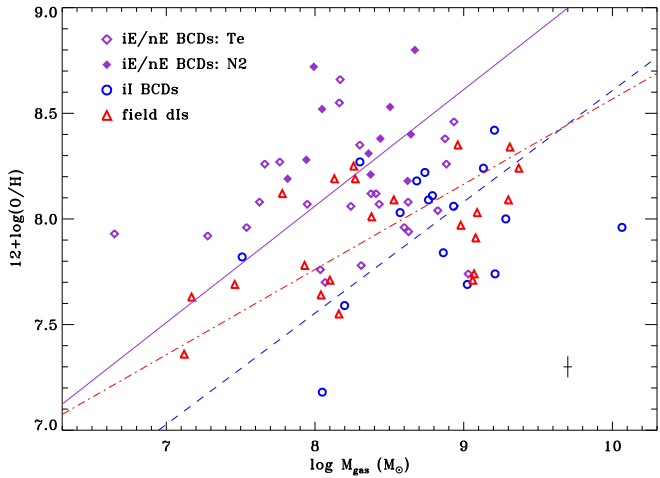}
\caption{Correlation between oxygen abundance and gas mass. The bottom right cross shows the typical error bar of the BCD sample. Albeit there is a large scatter, a galaxy with a greater gas mass is generally more abundant in metals. The slopes of this relation for I- and E-type BCDs are similar, and are steeper than that of field dIs. However, in general, E-type BCDs have a higher mean metallicity than I-type ones at a given gas mass.}
\label{Fig6}
\end{figure}

Although the slopes of the $M_{\rm gas}-Z$ relations for I- and E-type BCDs are almost the same, and there are larger scatters in these relations, it is clear that E-type BCDs have higher mean oxygen abundance than I-type ones at a given gas mass, suggesting that they are depleted in gas relative to I-type BCDs of comparable metallicity. Field dIs have a shallower $M_{\rm gas}-Z$ relation than BCDs, but are mixed with BCDs in the plot. Combining this difference with the fact that I-type BCDs are more gas-rich than E-type BCDs (see Section 3.2), it might indicate a possible difference in potential for the progenitors and/or evolutionary phases of E- and I-type BCDs.

\subsubsection{Metallicity versus Baryonic Mass}
Figure 7 plots the oxygen abundance as a function of the total baryonic mass. For E-type BCDs, unlike Figure 6, this figure shows a much better correlation between metallicity and baryonic mass. A geometric linear fit to the E-type sample (36 members) gives the following relation
\[12+\log ({\rm O/H})=(3.59\pm0.61)+(0.53\pm0.11)\log M_{\rm tot}.\]
with a rms scatter of 0.25 dex in (O/H). The fitted relation is plotted with a solid line. 

\begin{figure}[t]
\centering
\includegraphics[bb=20 8 490 344,width=0.48\textwidth]{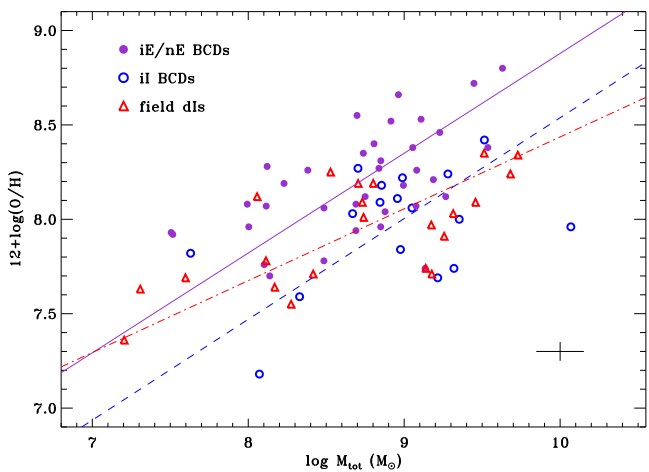}
\caption{Correlation between oxygen abundance and the total baryon mass $(M_{\rm{gas}}+M_\star)$, with the cross showing the typical error bar of the BCD sample. }
\label{Fig7}
\end{figure}

We also do the same fits to the I-type BCD and dI samples, which give
\[12+\log ({\rm O/H})=(3.20\pm1.22)+(0.53\pm0.18)\log M_{\rm tot}.\]
and
\[12+\log ({\rm O/H})=(4.63\pm0.33)+(0.38\pm0.06)\log M_{\rm tot}.\]
with rms scatters of 0.31 dex and 0.22 dex in (O/H), for the I-type BCDs and dIs, respectively. The dashed and dot-dashed lines superposed in Figure 7 are the respective fitting results for I-type BCDs and dIs.

These results show that the baryonic mass-metallicity relations for I- and E-type BCDs have similar slopes, which are a bit steeper than that of field dIs. However, E-type BCDs appear to lie at a higher oxygen abundance comparing with I-type ones at a given baryonic mass, which again seems to suggest that E- and I-type BCDs might be at different evolutionary phases and/or have different progenitors.

In fact, BCDs also have steeper \LZ\ and mass-metallicity (\MZ) relations than dIs (ZGG10; Zahid et al. 2011). Kunth \& \"{O}stlin (2000) show that there exist offsets between the $B$ band \LZ\ relations of BCDs, dIs and dEs/dSphs, in the sense that at a given $B$ luminosity it becomes relatively more metal-deficient following the dEs/dSphs, dIs and BCDs sequence. This offset persists even only comparing the metallicities of the old populations in dIs and dEs/dSphs (Grebel et al. 2003). Woo et al. (2008) find the same phenomenon in the \MZ\ relations of dIs and dEs/dSphs. In the current work, however, we find that dIs distribute in between the two types of BCDs in the baryonic-metallicity relation. Therefore, the evolutionary connections, if exist, between dEs, dIs, and BCDs may be complicated.

\subsubsection{Metallicity versus Gas Fraction}
The relation between metallicity and gas fraction can be used to check different chemical evolution models. The closed-box model predicts a linear correlation between the logarithm of the inverse gas fraction $1/\mu$ and the oxygen abundance by number, $n({\rm O})/n({\rm H})$, namely (e.g. Lee et al. 2003)
\begin{equation}
12+\log ({\rm O/H}) =11.3 +\log y_{\rm O} + \log \log (1/\mu)
\end{equation}
where $y_{\rm O}$ is the yield of oxygen and we have assumed that the fraction of gas in the form of hydrogen is $X=0.733$ (solar value). 

A comparison of the observed oxygen abundances to those predicted by closed-box models (thick solid line) is shown in Figure 8, which indicates that few of the galaxies in these BCD and dI samples agree with a closed-box chemical evolution. The vast majority of galaxies appear to have oxygen abundances less than that expected by closed-box models (with a Salpeter IMF), which imply varying IMF, and/or gas inflows/outflows in these galaxies. It is also interesting to find that I-type BCDs and dIs occupy a similar region in this plot, whereas E-type BCDs extend to the most metal-rich and gas-poor region. This indicates that (1) these galaxies follow a similar chemical evolution path (e.g. experiencing similar gas flows), but at different phase (e.g. E-type BCDs are more evolved than dIs and I-type BCDs); or (2) E-type BCDs experience different gas flows from dIs and I-type BCDs (see below).

The effects of inflows and/or outflows have been described by many authors (e.g., Edmunds 1990; Dalcanton 2007; Erb 2008;  Matteucci 2008; Spitoni et al. 2010). In particular, Dalcanton (2007) shows that inflows are very efficient in reducing the effective yield, $y_{eff}$, for gas-poor galaxies while outflows are more effective in changing $y_{eff}$ in gas-rich galaxies. Erb (2008) and Matteucci (2008) introduce a model in which galaxies have inflows/outflows proportional to the star formation rate (SFR). This model adopts the instance recycling and mixing approximations, and the infalling gas is assumed to have no metal, while the enriched outflows is considered to have the same metallicity as the interstellar medium (ISM). In this model, we can change two free parameters, the amount of outflow $f_o$ and inflow $f_i$ in unit of the SFR of the galaxy, to reproduce the observed oxygen abundance. Two more parameters can be varied, if necessary, i.e., the true yield $y$ (i.e. IMF) and the fraction $\alpha$ of mass still locked in stars.

In Figure 8 we also plot the predicted metallicity-gas fraction relations from this model using different parameters. We adopt $y_{\rm O}$ as 1 and $\frac{1}{5}$ of the true oxygen yield predicted by the closed-box model with a Salpeter IMF($y_{\rm O,\,true}=0.0074$; Meynet \& Maeder 2002), and vary $f_o$ and $f_i$ to reproduce the dependence of oxygen abundance on the inverse gas fraction. The parameter $\alpha$ is a slowly decreasing function of time, as more stars leave the main sequence, but can vary in quite a small range. It is expected to be $\sim 0.76$ at $t=10^9$ yr and $\sim0.69$ at $t=10^{10}$ yr (Bruzual \& Charlot 2003) for sub-solar metallicity galaxies with a Salpeter IMF. Given the age of these galaxies (see, e.g. ZGG11), we have used $\alpha=0.7$, in despite of that the results are not very sensitive to the chosen value.

For a Salpeter IMF, as shown in Figure 8, the pure inflow model (black dotted lines) with a variable rate can not explain the observed metallicity-gas fraction relation, (especially for E-type BCDs), which is consistent with the finding in Spitoni et al. (2010). Small inflow rate predicts a higher metallicity than the observed one, whereas high inflow rate can not reach the gas-poor region. Pure outflow model (thin solid lines) or the combination of outflow and inflow models (dashed and dot-dashed lines) can reproduce the observed distribution, but the pure outflow model may need very high flow rate. The best solution could be a combination of gas flows and a variable IMF. 

\begin{figure}[t]
\centering
\includegraphics[bb=20 7 490 350,width=0.48\textwidth]{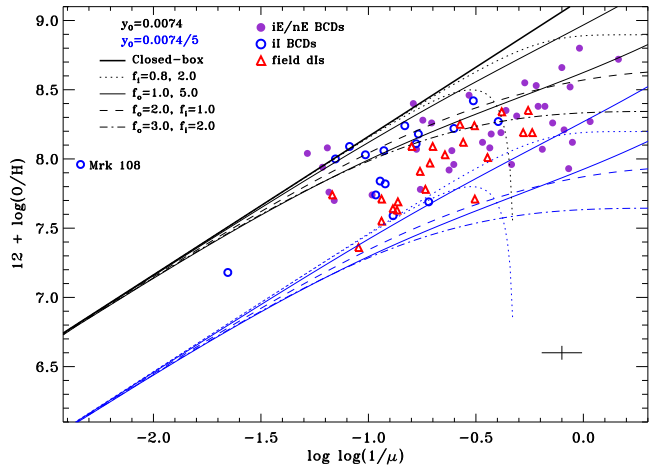}
\caption{Observed oxygen abundance plotted as a function of inverse gas fraction for E- and I-type BCDs and field dIs. The typical error bar of the BCD sample is indicated by the bottom right cross. Superposed lines are expectations from different models of chemical evolution. The thick solid line shows the result from a closed-box model with an oxygen yield approximate for low-metallicity galaxies with a Salpeter IMF ($y_{\rm O}=0.0074$; Meynet \& Maeder 2002). The thin solid (dashed) lines are pure outflow (inflow) models, with an effective yield 1 (black) and $\frac{1}{5}$ (blue) of the true yield. The rest (noncontinuous) lines show the results of applying the model in Erb (2008) with values of $f_o$, $f_i$ and $y_{\rm O}$ in the labels. The labeled galaxy, Mrk 108, which is an apparent merging system, appears to be interpreted by none of these models.}
\label{Fig8}
\end{figure}

\section{Discussion}
In this section we will mainly discuss the possible evolutionary connections between BCDs, dEs and dIs. There have been several evolutionary scenarios, both from observational and theoretical aspects, between BCDs, dEs, and dIs (see Papaderos et al. 1996 for a review). However, the question of whether there is an evolutionary connection between these three types of dwarf galaxies have not been achieved a solid answer yet. As mentioned in Section 1, some apparently contrary results are present in previous studies on the basis of observational measurements. 

For the BCDs in our sample, metallicity correlates with gas mass and total baryonic mass, in the sense that more massive systems are more abundant in metals. For these two correlations, I-type BCDs show similar slopes to E-type ones. However, I-type BCDs are generally more metal-poor and gas-rich than E-type ones at a given gas/baryonic mass. Although field dIs seem to have a shallower slope relative to BCDs, they are well mixed with BCDs in the distributions of $M_{\rm gas}-Z$ and $M_{\rm tot}-Z$ plots. Moreover, in the plot of metallicity-gas fraction relation, I-type BCDs generally occupy a region differing from E-type ones. 

Our results may shed some light on the evolutionary connection between the three types of dwarf galaxies. I- and E-type BCDs might be at different evolutionary phases and/or have different progenitors. For example, if some E-type BCDs are formed from dEs (e.g. Silk et al. 1987), whereas I-type ones are formed from dIs, then this part of E-type BCDs should be more evolved than I-type ones. This scenario is also supported by the fact that many E-type BCDs show structural properties compatible with dEs (Papaderos et al. 1996; Gil de Paz \& Madore 2005). In addition, field dIs show chemical and gaseous properties in between I- and E-type BCDs, which suggests that  they could be connected to both types of BCDs (e.g. Thuan 1985; Davies \& Phillipps 1988).

As discussed in Papaderos et al. (1996), the interchanges between dIs, BCDs and dEs need gas outflows/inflows to get involved, which is very consistent with our result that most BCDs should have experienced gas flows during their lives. Indeed, observations have provided (direct) evidence for gas flows in BCDs (e.g., Papaderos et al. 1994; Martin 1996, 1999; Martin et al. 2001; Elson et al. 2009; van Eymeren et al. 2009; L\'{o}pez-S\'{a}chez et al. 2010). Using X-ray data, Papaderos et al. (1994) estimate a galactic wind flowing at a velocity of $>1000$ km s$^{-1}$ for the BCD VII Zw 403, whereas the estimated escape velocity for this galaxy is only 50  km~ s$^{-1}$. Martin et al. (2001) have shown that the galactic wind in NGC 1569 carries away almost all the oxygen produced in the current starburst. More recently, Elson et al. (2009) present their \HI\ observations for NGC 2915, and find a gas component clearly lagging in velocity relative to the main gas disk of the galaxy. This might be evidence for gas accretion from inter-galactic space onto the outer galaxy disk. Van Eymeren et al. (2009) report their detailed kinematic studies on the neutral (\HI) and ionized (H$\alpha$) gas components in a nearby BCD, NGC 4861. The authors find that both in \HI\ and H$\alpha$ the galaxy shows several outflows. Using multi-wavelength data, another work by this group (L\'{o}pez-S\'{a}chez et al. 2010) have studied the environment of a sample of nearby BCDs, and shown that all BCDs present evident interacting features in their \HI\ component despite the environment in which they reside. In particular, IC 4870, which has an elliptical outer component, is accreting \HI\ clouds. 

\section{Summary}
We have presented a detailed studies on the gaseous and chemical properties of a sample of 53 BCDs, which is selected from G03/ZGG10 sample. According to their morphological information, this sample of BCDs is divided into two subsamples: E- and I-type BCD samples. An E-type BCD contains an elliptical outer component while an I-type BCD has an irregular outer halo.

For both types of BCDs, various correlations among metallicity, gaseous, stellar and baryonic mass, and gas fraction, are investigated. We find that in general, (1) more massive systems are more metal-rich, (2) a galaxy having a larger stellar mass also contains more gas, but a smaller gas fraction, (3) I- and E-type BCDs have similar slopes of the $M_{\rm gas}-Z$ and $M_{\rm tot}-Z$ relations, but I-type BCDs are generally more gas-rich and metal-poor. Based on the comparisons of these correlations, we suggest that I- and (at least a part of) E-type BCDs might be likely at different evolutionary stages and/or having progenitors. By comparing the metallicity-gas fraction relation for our sample with those predicted by chemical evolution models, we conclude that most BCDs should have experienced  some gas inflows/outflows and/or mergers. 

\begin{acknowledgements}
The authors gratefully acknowledge the anonymous referee for his/her constructive comments and suggestions that improved this paper. Research for this project is supported by the National Natural Science Foundation of China under grants 10833006 and 10903029. This research has made use of the NASA/IPAC Extragalactic Database (NED), which is operated by the Jet Propulsion Laboratory, California Institute of Technology, under contract with the National Aeronautics and Space Administration.\end{acknowledgements}

%\end{CJK*}

\begin{thebibliography}{}
\bibitem[]{}Allende Prieto, C., Lambert, D. L., \& Asplund, M. 2001, \apj, 556, L63
\bibitem[]{}Amor\'in, R., Alfonso, J., Aguerri, J. A. L., Mu\~noz-Tu\~n\'on, C., \& Cair\'os, L. M. 2009, A\&A, 501, 75
\bibitem[]{}Asplund, M., Grevesse, N., Sauval, A. J., \& Scott, P. 2009, \araa, 47, 481
\bibitem[]{}Bell, E. F., \& de Jong, R. F. 2001, \apj, 550, 212
\bibitem[]{}Binggeli, B. 1994, in Dwarf Galaxies, ed. G. Meylan \& P. Prugniel (Garching: ESO), 13
\bibitem[]{}Bruzual, G., \& Charlot, S. 2003, \mnras, 344, 1000
\bibitem[]{}Calzetti, D. 2001, PASP, 113, 1449
\bibitem[]{}Cannon, J. M., Skillman, E. D., Garnett, D. R., \& Dufour, R. J. 2002, \apj, 565, 931
\bibitem[]{}Cardelli, J. A., Clayton, G. C., \& Mathis, J. S. 1989, \apj, 345, 245
\bibitem[]{}Cair\'os, L. M., Caon, N., V\'ilchez, J. M., Gonz\'alez-P\'erez, J. N., \& Mu\~noz-Tu\~n\'on,  C. 2001a, ApJS, 136,  393
\bibitem[]{}Cair\'os, L. M., V\'ilchez, J. M., Gonz\'alez-P\'erez, J. N., Iglesias-P\'aramo, J., \& Caon, N. 2001b, ApJS, 133, 321
\bibitem[]{}Cair\'{o}s, L. M., Caon, N., Zurita, C., Kehrig, C., Roth, M., \& Weilbacher, P. 2010, \aap, 520, 90
\bibitem[]{}Cid Fernandes, R., Asari, N. V., Sodr\'e, L., Stasi\'nska, G., Mateus, A., Torres-Papaqui, J. P., \& Schoenell, W. 2007, MNRAS, 375, 16
\bibitem[]{} Cid Fernandes, R., Mateus, A., Sodr\'e, Jr, L., Stasi\'nska, G., \& Gomes, J. M. 2005, MNRAS, 358, 363
%\bibitem[]{}Conselice, C. J., O'Neil, K., Gallagher, J. S., \& Wyse, R. F. G. 2003, \apj, 591, 167
\bibitem[]{}Dalcanton, J. J. 2007, \apj, 658, 941
\bibitem[]{}Davies, J. I., \& Phillipps, S. 1988, \mnras, 233, 553
\bibitem[]{}Denicol\'o, G., Terlevich, R., \& Terlevich, E. 2002, \mnras, 330, 69
\bibitem[]{}Edmunds, M. G. 1990, \mnras, 246, 678
\bibitem[]{}Elson, E., de Blok, E., Kraan-Korteweg, R. C. 2009 in Proceedings of Panoramic Radio Astronomy: Wide-field 1-2 GHz research on galaxy evolution, ed. G. Heald \& P. Serra, 51
\bibitem[]{}Erb, D. K. 2008, \apj, 674, 151
\bibitem[]{}Ferguson, H. C., \& Binggeli, B. 1994, A\&A Rev., 6, 67
\bibitem[]{}Freedman, W. L., et al. 2001, \apj, 553, 47
\bibitem[]{}Gallazzi, A., \& Bell, E. F. 2009, \apjs, 185, 253
\bibitem[]{}Garc\'{i}a-Lorenzo, B., Cair\'{o}s, L. M., Caon, N., Monreal-Ibero, A., \& Kehrig, C. 2008, \apj, 677, 201
\bibitem[]{}Gavazzi, G., Boselli, A., van Driel, W., \& K. O'Neil 2005, A\&A, 429, 439
\bibitem[]{}Gil de Paz, A., Madore, B. F., \& Pevunova, O. 2003, \apjs, 147, 29 (G03)
\bibitem[]{}Gil de Paz, A., \& Madore, B. F. 2005, \apjs, 156, 345
\bibitem[]{}Gordon, D., \& Gottesman, S. T. 1981, \aj, 86, 161
\bibitem[]{}Grebel, E. K., Gallagher, III, J. S.,  \& Harbeck, D. 2003, \aj, 125, 1926
\bibitem[]{}Hoffman, G. L., Salpeter, E. E., Farhat, B., Roos, T., Williams, H., \& Helou, G. 1996, \apjs, 105, 269
\bibitem[]{}Huchtmeier, W. K., Krishna, G., \& Petrosian, A. 2005, A\&A, 434, 887
\bibitem[]{}Huchtmeier, W. K., \& Richter, O.-G. 1988, A\&A, 203, 237
\bibitem[]{}Hunter, D. A. 1997, PASP, 109, 937
\bibitem[]{}Hunter, D. A., \& Hoffman, L. 1999, \aj, 117, 2789
\bibitem[]{}Isobe, T., Feigelson, E. D., Akritas, M. G., \& Babu, G. J. 1990, \apj, 364, 104
\bibitem[]{}Izotov, Y. I., \& Thuan, T. X. 1999, \apj, 511, 639
\bibitem[]{}Izotov, Y. I., Thuan, T. X., \& Lipovetsky, V. A. 1994, \apj, 435, 647
\bibitem[]{}James, P. A. 1994, \mnras, 269, 176
\bibitem[]{} Kauffmann, G.,  White, S. D. M., \& Guiderdoni, B. 1993, MNRAS, 264, 201
\bibitem[]{} Kobulnicky, H. A., \& Skillman, E. D. 1996, \apj, 471, 211
\bibitem[]{} Kobulnicky, H. A., \& Skillman, E. D. 1997, \apj, 4789, 636
\bibitem[]{}Kunth, D., \& \"{O}stlin, G. 2000, A\&ARv, 10, 1
\bibitem[]{}Le Borgne, J.-F., et al. 2003, A\&A, 402, 433
\bibitem[]{}Lee, H., McCall, M. L., Kingsburgh, R. L., Ross, R., \& Stevenson, C. C. 2003, \aj, 125, 146
\bibitem[]{}Lee, H., \& Skillman, E. D. 2004, \apj, 614, 698
\bibitem[]{}Loose, H.-H., \& Thuan, T. X. 1986, in Star Forming Dwarf Galaxies and Related Objects, ed. D. Kunth, T. X. Thuan, \& J. T. T. Van (Gif-sur-Yvette: Editions Fronti\'{e}res), 73
\bibitem[]{}L\'{o}pez-S\'{a}nchez, A. R., Koribalski, B., van Eymeren, J., Esteban, C., Popping, A., \& Hibbard, J. 2010, in Galaxies in Isolation: Exploring Nature Versus Nurture, ed. L. Verdes-Montenegro, A. del Olmo, \& Jack Sulentic (San Francisco: Astronomical Society of the Pacific), 65
\bibitem[]{}Martin, C. L. 1996, \apj, 465, 680
\bibitem[]{}Martin, C. L. 1999, \apj, 513, 156
\bibitem[]{}Martin, C. L., Kobulnicky, H. A., \& Heckman, T. M. 2002, \apj, 574, 663
\bibitem[]{}Mateo, M. 1998, \araa, 36, 435
\bibitem[]{}Matteucci, F. 2008, arXiv: 0804.1492
\bibitem[]{}McCall, M. L., Vaduvescu, O., Pozo Nunez, F., Barr Dominguez, A., Fingerhut, R., Unda-Sanzana, E., Li, B., \& Albrecht, M. 2012, A\&A, 540, 49
\bibitem[]{}Meynet, G., \& Maeder, A. 2002. A\&A, 390, 561
\bibitem[]{}Moustakas, J., \& Kennicutt, R. C., Jr. 2006, ApJS, 164, 81
\bibitem[]{}Noeske, K. G., Guseva, N. G., Fricke, K. J., Izotov, Y. I., Papaderos, P., \& Thuan, T. X. 2000, A\&A, 361, 33
\bibitem[]{}Noeske, K. G., Papaderos, P., Cair\'os, L. M., \& Fricke, K. J. 2003, A\&A, 410, 481
\bibitem[]{}Papaderos, P., Fricke, K. J., Thuan, T. X., \& Loose, H.-H. 1994, A\&A, 291, L13
\bibitem[]{}Papaderos, P., Loose, H.-H., Fricke, K. J., \& Thuan, T. X. 1996, A\&A, 314, 59
\bibitem[]{}Pettini, M., \& Pagel, B. E. J. 2004, \aj, 348, L59
\bibitem[]{}Richer, M. G., \& McCall, M. L. 1995, \apj, 445, 642
\bibitem[]{}Roberts, M. S. 1975, in Galaxies and the Universe, ed. A. Sandage, M. Sandage, \& J. Kristian (Chicago: Univ. Chicago Press), 309 
\bibitem[]{}Roberts, M. S., \& Haynes, M. P. 1994, \araa, 32, 115
\bibitem[]{}Salpeter, E. E. 1955, \apj, 121, 161
\bibitem[]{}Salzer, J. J., Lee, J. C., Melbourne, J., Hinz, J. L., Alonso-Herrero, A., \& Jangren, A. 2005, \apj, 624, 661
\bibitem[]{}Salzer, J. J., MacAlpine, G. M., \& Boroson, T. A. 1989, 70, 479
\bibitem[]{}Schlegel, D. J., Finkbeiner, D. P., \& Davis, M. 1998, \apj, 500, 525
\bibitem[]{}Schulte-Ladbeck, R. E., Hopp, U., Greggio, L., \& Crone, M. M. 2000, \aj, 120, 1713
\bibitem[]{}Silk, J., Wyse, R. F. G., \& Shields, G. A. 1987, \apj, 322, L59
\bibitem[]{}Spitoni, E., Calura, F., Matteucci, F., \& Recchi, S 2010, A\&A, 514, 73
\bibitem[]{}Telles, E., Melnick, J., \& Terlevich, R. 1997, \mnras, 288, 78
\bibitem[]{}Terlevich, R., Melnick, J., Masegosa, J., Moles, M., \& Copetti, M. V. F. 1991, A\&AS, 91, 285
\bibitem[]{}Thuan, T. X. 1985, \apj, 299, 881
\bibitem[]{}Thuan, T. X., Lipovetsky, V. A., Martin, J.-M., \& Pustilnik, S. A. 1999, A\&AS, 139, 1
\bibitem[]{}Thuan, T. X., \& Martin, G. E. 1981, \apj, 247, 823
\bibitem[]{}Tolstoy, E., Hill, V., \& Tosi, M. 2009, \araa, 47, 371
\bibitem[]{}Vaduvescu, O., \& McCall, M. L. 2008, A\&A, 487, 147
\bibitem[]{}Vaduvescu, O., McCall, M. L., \&  Richer, M. G., 2007, \aj, 134, 604 (VMR07)
\bibitem[]{}Vaduvescu, O., Richer, M. G., \& McCall, M. L. 2006, \aj, 131, 1318
\bibitem[]{}van Eymeren, J., Marcelin, M., Koribalski, B. S., Dettmar, R.-J., Bomans, D. J., Gach, J.-L., \& Balard, P. 2009, A\&A, 505, 105
\bibitem[]{}Walcher, C. J., et al. 2008, A\&A, 491, 713
\bibitem[]{}Woo, J., Courteau, S., \& Dekel, A. 2008, \mnras, 390, 1453
\bibitem[]{}Zahid, H. J., Bresolin, F., Kewley, L. J., Coil, A. L., \& Dav\'{e}, R. 2012, \apj, 750, 120
\bibitem[]{}Zhao, Y., Gao, Y., \& Gu, Q. 2010, \apj, 710, 663 (ZGG10)
\bibitem[]{}Zhao, Y., Gu, Q., \& Gao, Y. 2011, \aj, 141, 68 (ZGG11)
\bibitem[]{}Zwicky, F. 1966, \apj, 143, 192
%%%%----------

\end{thebibliography}
\end{document}